\documentclass{emulateapj}
\usepackage{natbib}
\usepackage{apjfonts}
\begin{document}

\title{A $5 \times 10^9\ M_\odot$ Black Hole in NGC 1277 from Adaptive
  Optics Spectroscopy}

\author{Jonelle L. Walsh$^{1}$, Remco C.~E. van den Bosch$^{2}$, Karl
  Gebhardt$^{3}$, Ak{\i}n Y{\i}ld{\i}r{\i}m$^{2}$, Douglas
  O. Richstone$^{4}$, Kayhan G\"{u}ltekin$^{4}$, {\sc and} Bernd
  Husemann$^{5}$}

\affil{$^1$ George P. and Cynthia Woods Mitchell Institute for
  Fundamental Physics and Astronomy, Department of Physics and
  Astronomy, Texas A\&M University, College Station, TX 77843, USA;
  walsh@physics.tamu.edu \\
  $^2$ Max-Planck Institut f\"{u}r Astronomie, K\"{o}nigstuhl 17,
  D-69117 Heidelberg, Germany\\
  $^3$ Department of Astronomy, The University of Texas at Austin,
  2515 Speedway, Stop C1400, Austin, TX 78712, USA \\
  $^4$ Department of Astronomy, University of Michigan, 1085
  S.~University Ave., Ann Arbor, MI 48109, USA\\
  $^5$ European Southern Observatory, Karl-Schwarzschild-Str. 2, 85748
  Garching, Germany}

\begin{abstract}

  The nearby lenticular galaxy NGC 1277 is thought to host one of the
  largest black holes known, however the black hole mass measurement
  is based on low spatial resolution spectroscopy. In this paper, we
  present Gemini Near-infrared Integral Field Spectrometer
  observations assisted by adaptive optics. We map out the galaxy's
  stellar kinematics within $\sim$440 pc of the nucleus with an
  angular resolution that allows us to probe well within the region
  where the potential from the black hole dominates. We find that the
  stellar velocity dispersion rises dramatically, reaching $\sim$550
  km s$^{-1}$ at the center. Through orbit-based, stellar-dynamical
  models we obtain a black hole mass of $(4.9\pm1.6) \times 10^9\
  M_\odot$ (1$\sigma$ uncertainties). Although the black hole mass
  measurement is smaller by a factor of $\sim$3 compared to previous
  claims based on large-scale kinematics, NGC 1277 does indeed contain
  one of the most massive black holes detected to date, and the black
  hole mass is an order of magnitude larger than expectations from the
  empirical relation between black hole mass and galaxy
  luminosity. Given the galaxy's similarities to the higher redshift
  ($z$$\sim$2) massive quiescent galaxies, NGC 1277 could be a relic,
  passively evolving since that period. A population of local analogs
  to the higher redshift quiescent galaxies that also contain
  over-massive black holes may suggest that black hole growth precedes
  that of the host galaxy.

\end{abstract}


\section{Introduction}
\label{sec:intro}

The relationships between black hole mass, $M_\mathrm{BH}$, and
large-scale host galaxy properties, such as bulge luminosity,
$L_\mathrm{bul}$, and stellar velocity dispersion, $\sigma_\star$
(e.g., \citealt{Kormendy_Richstone_1995, Ferrarese_Merritt_2000,
  Gebhardt_2000, Marconi_Hunt_2003, Gultekin_2009}) suggest that black
holes play a fundamental role in the evolution of galaxies. It is
believed that supermassive black holes and galaxies grow in tandem,
with the black hole governing the properties of galaxies through
feedback mechanisms \citep{Silk_Rees_1998, Fabian_1999}. However, the
scaling relations can also be established without black hole feedback,
simply owing to the inherent averaging of properties that accompanies
galaxy mergers \citep{Peng_2007, Jahnke_Maccio_2011}. The main
physical process that drives the black hole-galaxy relations remains
poorly understood and it is unclear how important feedback-driven
co-evolution is compared to a non-causal link associated with merger
averaging (e.g., \citealt{Kormendy_Ho_2013}). A better understanding
of the role that black holes play in galaxy evolution requires
increasing the number of $M_\mathrm{BH}$ measurements, especially for
low and high-mass black holes and over a larger range of galaxy
types/properties.

NGC 1277 is thought to harbor one of the most massive black holes in
the Universe \citep{vandenBosch_2012}. The nearby lenticular galaxy is
a flattened, fast rotator with a cuspy surface brightness profile. Of
the $\sim$80 dynamical black hole mass measurements currently on the
black hole-galaxy relations \citep{McConnell_Ma_2013,
  Kormendy_Ho_2013}, NGC 1277 shares the most similarities with NGC
4342 \citep{Cretton_vandenBosch_1999}, NGC 1332 \citep{Rusli_2011},
and NGC 1271 \citep{Walsh_2015}. All of these early-type galaxies are
significant positive outliers from the $M_\mathrm{BH} -
L_\mathrm{bul}$ relation (by at least 2$\sigma$), and have small sizes
(effective radii, $R_\mathrm{e}$, below 3 kpc) and large stellar
velocity dispersions ($225-333$ km s$^{-1}$) for their $K$-band
luminosities ($2.7\times10^{10}- 1.6\times10^{11}\ L_\odot$). NGC
4486B \citep{Kormendy_1997} and M60-UCD1 \citep{Seth_2014} are also
extreme outliers from $M_\mathrm{BH} - L_\mathrm{bul}$ with small
sizes and elevated velocity dispersions for their galaxy luminosities.

The compact, high-dispersion galaxies like NGC 1277 are distinct from
the typical giant elliptical galaxies and Brightest Cluster Galaxies
(BCGs) that are expected to contain the largest black holes and occupy
the upper end of the black hole-galaxy relationships. Giant
ellipticals and BCGs instead have large sizes, surface brightness
profiles with flat central cores, are pressure supported, and often
are slow rotators \citep{DallaBonta_2009, McConnell_2012}. Therefore
measuring black hole masses in the compact, high-dispersion galaxies
and the giant ellipticals/BCGs is helpful for searching for systematic
deviations in the black hole scaling relationships between the two
kinds of early-type galaxies. If such differences arise, this would
suggest that the black holes and galaxies grew in different manners.

Galaxies like NGC 1277, however, appear remarkably similar to the
higher redshift massive quiescent galaxies. The $z\sim$2 quiescent
galaxies are also small, have large velocity dispersions, exhibit
disky structures (e.g., \citealt{Daddi_2005, Trujillo_2006,
  vanDokkum_2009, vanderWel_2011}), and are thought to grow into the
present-day massive galaxies through a series of mergers (e.g.,
\citealt{vanDokkum_2010, Wellons_2015}). Given the resemblance to the
$z\sim$2 red nuggets, the compact, high-dispersion galaxies could be
relics that followed a different pathway than the local massive
galaxies, instead experiencing passive evolution \citep{Trujillo_2014,
  FerreMateu_2015}. If galaxies like NGC 1277 are local analogs of the
$z\sim$2 quiescent galaxies and continue to show evidence for
harboring over-massive black holes, then such galaxies may reflect a
previous period when the local black hole scaling relations did not
apply. Instead, galaxies may have contained over-massive black holes
at higher redshifts, suggesting that black hole growth precedes that
of its host galaxy. Indeed there could be hints that black holes at
higher redshifts (masses estimated in active galactic nuclei through
the width of broad emission lines) are over-massive at fixed galaxy
mass compared to the local relation (e.g., \citealt{Peng_2006,
  Woo_2008, Merloni_2010, Decarli_2010, Bennert_2011,
  Trakhtenbrot_2015}), however there are also a number of studies that
find essentially no change in the black hole scaling relations with
redshift (e.g., \citealt{Salviander_2013, Salviander_2015, Shen_2015})
and selection bias can lead to the false identification of evolution
in $M_\mathrm{BH} - \sigma_\star$ and $M_\mathrm{BH} - L_\mathrm{bul}$
\citep{Lauer_2007, Schulze_2014}.

The black hole in NGC 1277 has several previous $M_\mathrm{BH}$
measurements. \cite{vandenBosch_2012} first reported the presence of a
$1.7 \times 10^{10}\ M_\odot$ black hole using stellar-dynamical
models and a long-slit spectroscopic observation from the Hobby-Eberly
Telescope (HET) at McDonald Observatory. The total stellar mass of NGC
1277 is $1.2 \times 10^{11}\ M_\odot$, and therefore the ratio between
black hole mass and galaxy stellar mass is surprisingly high at
14\%. \cite{vandenBosch_2012} also listed another five galaxies found
through the HET Massive Galaxy Survey \citep{vandenBosch_2015} that
could host similarly large black holes. Such an extraordinary black
hole mass for NGC 1277, however, was called into question by
\cite{Emsellem_2013}, who compared dynamical models from N-body
realizations to the same HET kinematics. \cite{Emsellem_2013} did not
search for a best-fit model, but demonstrated that a $5 \times 10^9\
M_\odot$ black hole was sufficient to reproduce the observed
kinematics, with the exception of the central $h_4$ measurements. More
recently, \cite{Yildirim_2015} used integral-field unit (IFU)
observations from the 3.5m telescope at Calar Alto Observatory and
orbit-based dynamical models to confirm the mass measurement of
\cite{vandenBosch_2012}, although they extend the bound of the
uncertainty on the black hole mass measurement to possibly as low as
$4.0 \times 10^9\ M_\odot$.

While the above black hole mass determinations were made from
analyzing the stellar kinematics of NGC 1277, gas kinematics can also
be used constrain $M_\mathrm{BH}$. \cite{Scharwachter_2015} presented
the detection of cold molecular gas arising from the nuclear dust disk
based on IRAM Plateau de Bure Interferometer (PdBI) observations. The
CO(1-0) emission was spatially unresolved, but the double-horned
profile suggested an enclosed mass of $\sim$$2 \times 10^{10}\
M_\odot$. Depending on the stellar mass-to-light ratio assumed, both a
$\sim$$1.7 \times 10^{10}\ M_\odot$ or a $\sim$$5 \times 10^9\
M_\odot$ black hole is able match the observed CO(1-0) kinematics.

Resolving the uncertainties in the NGC 1277 black hole mass requires
high angular resolution observations that probe well within the region
where the potential from the black hole dominates (the black hole
sphere of influence, $GM_\mathrm{BH}/\sigma_\star^2$). Here we present
new IFU observations from the 8m Gemini North telescope assisted by
laser guide star adaptive optics (AO) and use orbit-based stellar
dynamical models to revisit the black hole mass measurement of NGC
1277. In Section \ref{sec:mge}, we review the luminous mass model
adopted for NGC 1277. We describe the AO spectroscopic observations in
Section \ref{sec:nifs_obs}, the measurement of the high-angular
resolution stellar kinematics in Section \ref{sec:nifskin}, and the
corresponding PSF model in Section \ref{sec:psf}. Although the
kinematics measured from the AO observations are used to constrain
$M_\mathrm{BH}$ in our final model, we also run models that
simultaneously fit to large-scale kinematics extending to several
galaxy effective radii. These large-scale kinematics are discussed in
Section \ref{sec:largescalespectra}. We describe the dynamical models
in Section \ref{sec:modeling} and the results are given in Section
\ref{sec:results}. A discussion of the revised $M_\mathrm{BH}$ in
comparison to past work and implications from the location of NGC 1277
on the black hole-galaxy relations is provided in Section
\ref{sec:discussion}, followed by concluding remarks in Section
\ref{sec:conclusion}.

Throughout the paper, we assume a distance of 71 Mpc, which is the
Hubble flow distance from the Virgo + Great Attractor + Shapley
Supercluster Infall model \citep{Mould_2000} for $H_0 = 70.5$ km
s$^{-1}$ Mpc$^{-1}$, $\Omega_M = 0.27$ and $\Omega_\Lambda = 0.73$. At
this distance, 1\arcsec\ corresponds to 0.34 kpc. We further adopt
$R_\mathrm{e} = $ 3\farcs5 (1.2 kpc; \citealt{Yildirim_2015}) and
$\sigma_\star = 333$ km s$^{-1}$ \citep{vandenBosch_2012,
  Kormendy_Ho_2013}, which is also in agreement with the effective
stellar velocity dispersion from large-scale IFU observations reported
by \cite{Yildirim_2015}. For the galaxy's bulge luminosity, we apply
bulge-to-total ($B/T$) ratios determined through photometric
bulge-disk decompositions of a \emph{Hubble Space Telescope}
(\emph{HST}) $V$-band image to the 2MASS $K_s$ total luminosity
\citep{Jarrett_2000}. However, we note that a dynamical decomposition
derived from the best-fit model of \cite{Yildirim_2015} shows a
rotating central round component, in addition to several other
strongly rotating elements, making NGC 1277's dynamical structure
fundamentally different from BCGs and complicating the identification
of a bulge. Therefore, NGC 1277's bulge luminosity is uncertain and we
conservatively use $L_\mathrm{bul} = (7.7^{+2.8}_{-4.3}) \times
10^{10}\ L_\odot$ in the $K$-band, which comes from a bulge-to-total
ratio of $0.55$ \citep{Kormendy_Ho_2013}. The bounds on the bulge
luminosity are taken from the photometric decomposition of
\cite{Yildirim_2015}, who find lower and upper limits of $B/T = 0.24$
and $B/T = 0.75$. This is also consistent with a $B/T = 0.27$ measured
by \cite{vandenBosch_2012}.

\section{Luminous Mass Model}
\label{sec:mge}

We used the two-dimensional (2D) light distribution presented in
\cite{vandenBosch_2012}, which was constructed from an archival
\emph{HST} Advanced Camera for Surveys (ACS) image. The ACS data were
originally acquired under program GO-10546 with the F550M filter. The
observations have a total integration time of 2439 s, and a spatial
scale of 0\farcs05 pixel$^{-1}$. The light distribution is described
using the Multi-Gaussian Expansion (MGE) method \citep{Monnet_1992,
  Emsellem_1994}, as such a representation can be analytically
deprojected to determine the galaxy's intrinsic luminosity
density. \cite{vandenBosch_2012} generated the NGC 1277 MGE using
Galfit \citep{Peng_2010} while accounting for a Tiny Tim PSF model
\citep{Krist_Hook_2004} that was dithered in the same manner as the
galaxy observations. The elliptical galaxy NGC 1278, located
47\arcsec\ away in projection, was simultaneously fit with NGC 1277,
however other sources, including a nuclear dust disk, were masked and
excluded from the fit. The dust disk has an apparent axis ratio of
0.3, suggesting that the galaxy is viewed near edge-on, with an
inclination angle of 75$^\circ$. As presented in Supplemental Table 1
of \cite{vandenBosch_2012}, the NGC 1277 MGE consists of 10
components, each with the same center and position angle of
92.7$^\circ$. The dispersion of the Gaussians along the major axis
extends from 0\farcs06 to 17\arcsec, and the apparent flattening
ranges between 0.4-0.9.

\section{AO NIFS Observations}
\label{sec:nifs_obs}

We obtained high angular resolution spectroscopy of NGC 1277 using the
Near-infrared Integral Field Spectrometer (NIFS;
\citealt{McGregor_2003}) with the ALTtitude conjugate Adaptive optics
for the InfraRed \citep{Herriot_2000, Boccas_2006} system on the
Gemini North telescope. The observations were taken as part of program
GN-2011B-Q-27 over the course of four nights, spanning from 2012 Oct
30 to 2012 Dec 27. We observed NGC 1277 using 600 s Object-Sky-Object
exposures with the $H+K$ filter and $K$ grating centered on $2.2$
$\mu$m. For the galaxy observations we used the ``open-loop'' focus
model \citep{Krajnovic_2009} and the bright, compact nucleus as the
tip-tilt star. We further acquired NIFS observations of a nearby star
to measure the point spread function (PSF) and A0~V stars for telluric
correction.

We reduced the NIFS data using IRAF tasks\footnote[1]{IRAF is
  distributed by the National Optical Astronomy Observatory, which is
  operated by the Association of Universities for Research in
  Astronomy under cooperative agreement with the National Science
  Foundation} in the Gemini/NIFS package version 1.11. The reduction
followed the procedure outlined in the NIFS example processing
scripts\footnote[2]{http://www.gemini.edu/sciops/instruments/nifs/data-format-and-reduction}. The
main steps included sky subtraction, flat fielding, bad pixel and
cosmic-ray cleaning, spectral and spatial rectification, telluric
correction, and the creation of a data cube. Individual data cubes of
the galaxy were summed along the wavelength axis to generate flux
maps, which there then cross-correlated in order to determine the
relative spatial offsets between cubes. Twelve galaxy exposures,
totaling 2 hours-on source, were aligned and combined to produce the
final data cube composed of a spectral dimension and $x$ and $y$
spatial dimensions with a scale of 0\farcs05 pixel$^{-1}$. The PSF
observations were reduced in a similar manner.

We also retrieved NIFS $K$-band observations of K and M
giant/supergiant stars, and the supporting calibration files, from the
Gemini Science Archive. The twelve stars were originally observed
under programs GN-2006A-C-11, GN-2006B-Q-107, GN-2007A-Q-62, and
GN-2010A-Q-112. We followed the NIFS data reduction steps described
above, but additionally extracted a one-dimensional spectrum for each
star. We then rebinned the template spectra to the same wavelength
range and sampling, and shifted the stars to rest. Nearly all of the
stars are part of the NIFS Spectral Template Library
v2.0\footnote[3]{http://www.gemini.edu/sciops/instruments/nearir-resources/spectral-templates}
\citep{Winge_2009}, however we chose to reduce the observations from
scratch so that our template spectra would have a similar continuum
shape as the observed galaxy spectra. Currently, the spectra
associated with the NIFS Spectral Template Library have been
continuum-divided on a star-by-star basis.

\section{High Angular Resolution Stellar Kinematics}
\label{sec:nifskin}

From the NIFS data cube, we extracted spectra over a range of spatial
locations and used the Voronoi binning method
\citep{Cappellari_Copin_2003} to determine the optimal balance between
signal-to-noise (S/N) and bin size. We measured the line-of-sight
velocity distribution (LOSVD) in each spatial bin using the penalized
pixel fitting (pPXF) method \citep{Cappellari_Emsellem_2004},
parametrizing the LOSVD in terms of the radial velocity ($V$), the
velocity dispersion ($\sigma$), and the next two Gauss-Hermite moments
($h_3$, $h_4$) that correspond to asymmetric and symmetric deviations
from a Gaussian. An optimal stellar template in each spatial bin was
constructed from a library of twelve stars observed with NIFS. The
library includes K0--M5 giant stars, a K5 supergiant, and an M0
supergiant. Slight differences in the continuum shape and the
equivalent width between the LOSVD-broadened optimal template and the
observed galaxy spectrum were corrected for using an additive contant
and a first degree multiplicative Legendre polynomial.

We fit between observed wavelengths of $2.253 - 2.377$ $\mu$m, and
measured the kinematics primarily from the $(2-0)^{12}$CO and
$(3-1)^{12}$CO stellar features. We chose this particular wavelength
range to avoid artifacts in the outer spatial bins near the
$(4-2)^{12}$CO bandhead and to avoid the \ion{Na}{1} absorption line,
as were unable to achieve satisfactory fits due to the lack of dwarf
stars in our template library \citep{Krajnovic_2009, Walsh_2012}. For
the same reason, we excluded the \ion{Ca}{1} feature contained within
our fitting region. We further excluded an artifact near the
$(3-1)^{12}$CO bandhead.

Using the results of an initial fit to the observed galaxy spectrum in
each spatial bin, we ran Monte Carlo simulations with 100
realizations. During each realization, we added random Gaussian noise
to the best-fit model, such that the level of the perturbation is
given by the standard deviation of model residuals. We set the pPXF
penalization to zero to produce realistic uncertainties. From the
resulting distribution for each Gauss-Hermite moment in each of the
spatial bins, we calculated the mean and standard deviation, which we
took to be the kinematic value and 1$\sigma$ uncertainty.

We tested the robustness of the NIFS kinematics by modifying the
details of how the kinematics were extracted. Nearly all spatial bins
have kinematics consistent at the 1$\sigma$ level (all are consistent
within 2$\sigma$) with those measured using various combinations of
degree $0-1$ additive/multiplicative polynomials, as well as simply a
multiplicative order 1 polynomial. Similar kinematics are also
measured when including the \ion{Ca}{1} feature, and when fitting over
a longer wavelength region that extends to $2.415$ $\mu$m and includes
the third major CO bandhead. Moreover, we tested keeping the relative
mix of stars making up the optimal template fixed between spatial
bins, but allowing the coefficients of the additive and multiplicative
polynomial to vary. This produced consistent results as our default
method of fitting for the optimal stellar template in each spatial
bin.

\begin{figure}
\begin{center}
\epsscale{1.0}
\plotone{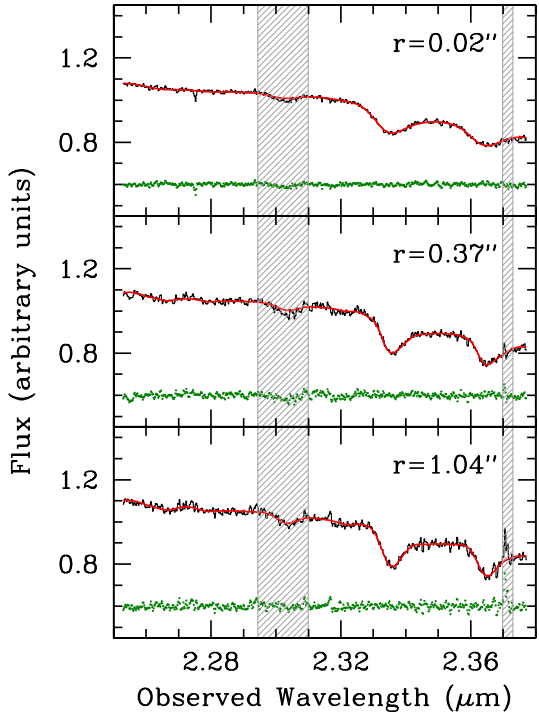}
\caption{Example fits to the NIFS $K$-band CO bandheads at the nucleus
  (top), at an intermediate distance from the galaxy center (middle),
  and in one of the outermost spatial bins (bottom). The best-fit
  stellar template broadened by the LOSVD is plotted in red, and the
  model residuals, after adding an arbitrary constant, are displayed
  in green. Gray shaded boxes indicate the wavelength regions excluded
  from the fit. \label{fig:nifsspec}}
\end{center}
\end{figure}

As a final step, we point-symmetrize the NIFS kinematics using the
procedure in \cite{vandenBosch_deZeeuw_2010}. We find that galaxy is
rotating with velocities of $\pm350$ km s$^{-1}$. The velocity
dispersion increases sharply to 551 km s$^{-1}$ at the center from
values of $\sim$275 km s$^{-1}$ in the outer NIFS spatial bins. As is
expected for an axisymmetric system, we find an anti-correlation
between the $h_3$ and $V$ maps. We also observe a slight peak in $h_4$
at the nucleus to values of $\sim$0.05. The median error for $V$,
$\sigma$, $h_3$, and $h_4$ over the 105 spatial bins is 7 km s$^{-1}$,
9 km s$^{-1}$, 0.02, and 0.02, respectively. Taking the S/N to be the
median flux divided by the standard deviation of model residuals from
the initial fits with pPXF, we find S/N values of 69$-$164, with a
median value of 92. Examples of the spectral fitting at three spatial
locations are shown in Figure \ref{fig:nifsspec}, and the kinematics
prior to symmetrization are given in Table \ref{tab:nifskin}.

\begin{deluxetable*}{rrrrrrrrrr}
\tabletypesize{\scriptsize}
\tablewidth{0pt}
\tablecaption{NIFS Kinematics \label{tab:nifskin}}
\tablehead{
\colhead{$x$ (\arcsec)} &
\colhead{$y$ (\arcsec)} &
\colhead{$V$ (km s$^{-1}$)} &
\colhead{$\Delta V$ (km s$^{-1}$)} &
\colhead{$\sigma$ (km s$^{-1}$)} &
\colhead{$\Delta \sigma$ (km s$^{-1}$)} &
\colhead{$h_3$} &
\colhead{$\Delta h_3$} &
\colhead{$h_4$} &
\colhead{$\Delta h_4$} \\
\colhead{(1)} &
\colhead{(2)} &
\colhead{(3)} &
\colhead{(4)} &
\colhead{(5)} &
\colhead{(6)} &
\colhead{(7)} &
\colhead{(8)} &
\colhead{(9)} &
\colhead{(10)}
}

\startdata

   0.004  &  -0.021  &    -8.361  &   9.999  &   579.913  &  15.292  &  -0.015  &   0.011  &   0.055  &   0.015  \\
   0.004  &   0.029  &   -21.369  &   7.795  &   522.633  &  11.262  &  -0.007  &   0.012  &   0.037  &   0.015  \\
  -0.046  &  -0.021  &   122.570  &   6.853  &   518.453  &   9.756  &  -0.035  &   0.010  &   0.061  &   0.013  \\
  -0.046  &   0.029  &    98.918  &   6.632  &   469.906  &   7.982  &  -0.030  &   0.013  &   0.039  &   0.014  \\
   0.054  &  -0.021  &   -44.814  &   7.965  &   575.108  &  17.143  &   0.007  &   0.011  &   0.061  &   0.016

\enddata

\tablecomments{The Voronoi bin $x$ and $y$ generators are given in
  Columns (1) and (2), and the NIFS unsymmetrized kinematics and
  errors are presented in Columns (3)-(10). The position angle is
  177.24$^\circ$ counterclockwise from the galaxy's major axis to
  $x$. This table is available in its entirety in machine-readable
  form.}

\end{deluxetable*}

\section{NIFS PSF Measurement}
\label{sec:psf}

We modeled the NIFS PSF as the sum of two concentric, circular 2D
Gaussians. The weights and dispersions of the components were
determined by comparing the galaxy data cube, after summing along the
wavelength axis, to the MGE model of the \emph{HST} image convolved
with the NIFS PSF model. The comparison was also used to determine the
center of the NIFS aperture. The NIFS PSF was best described by an
inner Gaussian with a dispersion of 0\farcs07 and a relative weight of
0.65 and an outer Gaussian with a dispersion of 0\farcs40 and a weight
of 0.35. Using more than two Gaussians to model the PSF produced
components with negligible weights.

Since the NIFS PSF is a vital input into the dynamical models when
measuring $M_\mathrm{BH}$, we also estimated the PSF by examining NIFS
observations of a nearby star. Using Galfit, we found that the sum of
four concentric, circular Gaussians provided a good description of the
collapsed data cube. The best-fit model was composed of Gaussians
weighted by 0.14, 0.31, 0.34, and 0.20 with dispersions of 0\farcs04,
0\farcs08, 0\farcs29, and 0\farcs72, respectively. Therefore, we
recover a similar NIFS PSF model using this alternative method, and
both NIFS PSFs are typical of the Gemini AO system
\citep[e.g.,][]{Krajnovic_2009, Gebhardt_2011, Onken_2014,
  Seth_2014}. Although we utilize the two Gaussian PSF model described
above for the final dynamical models, in Section \ref{sec:results} we
also test the effect on $M_\mathrm{BH}$ of adopting the four-component
model estimated from the NIFS PSF star observations.

\section{Large-scale Spectroscopy}
\label{sec:largescalespectra}

The NIFS kinematics are the primary data used to constrain the NGC
1277 black hole mass in Section \ref{sec:results}. However, we also
tested running models that include large-scale stellar kinematics
measured from HET Low Resolution Spectrograph (LRS;
\citealt{Hill_1998}) data and spectroscopy obtained at the 3.5m
telescope at Calar Alto Observatory using the Postdam Multi Aperture
Spectrograph (PMAS; \citealt{Roth_2005}) in the Pmas fiber PAcK (PPAK;
\citealt{Verheijen_2004, Kelz_2006}) mode. The HET and PPAK
observations, stellar kinematics (measured out to $\sim$3
$R_\mathrm{e}$), and PSF models have been previously presented by
\cite{vandenBosch_2012} and \cite{Yildirim_2015}. Although we refer
the reader to these publications for additional details, we briefly
review the relevant information below.

The HET observations were taken with a 1\arcsec-wide slit, the g2
grating, and 2$\times$2 binning, which provides an instrumental
dispersion of 108 km s$^{-1}$ and a wavelength coverage of $4200 -
7400$ \AA. A single 900 s exposure was obtained along the galaxy major
axis. After constructing 31 spatial bins, the stellar kinematics were
measured with pPXF using the the MILES template library
\citep{SanchezBlazquez_2006, FalconBarroso_2011} and a degree 25
multiplicative polynomial and an order 5 additive polynomial. By
comparing the reconstructed slit image to the NGC 1277 MGE model and
parameterizing the PSF as the sum of two Gaussians, weights of 0.66
and 0.34 and dispersions of 0\farcs62 and 2\farcs80, respectively,
were measured.

For the PPAK observations, two 900 s science exposures at three
pointings (totaling 1.5 hours on-source) were acquired using the
medium-resolution V1200 grating. With this setup, the useful spectral
range is $3650 - 4620$ \AA\ and the spectral resolution corresponds to
$R\sim1650$ at $4000$ \AA. The PPAK kinematic measurements for NGC
1277 were made within 38 Voronoi spatial bins using pPXF, with an
additive degree 15 Legendre polynomial, and the Indo-U.S. Library of
Coud\'{e} Feed Stellar Spectra \citep{Valdes_2004}. The PPAK PSF is
described by the sum of two Gaussians with dispersions of 1\farcs21
and 2\farcs44 with weights of 0.75 and 0.25, respectively. Thus, the
PPAK data have a lower spatial resolution than the HET observations.

\section{Dynamical Models}
\label{sec:modeling}

We constrained the mass distribution of NGC 1277 using the
three-integral, triaxial, orbit-based models from
\cite{vandenBosch_2008}. The black hole mass recovery of the code was
previously tested by \cite{vandenBosch_deZeeuw_2010}. With this
method, the galaxy's gravitational potential is constructed from the
combination of the black hole, the stars, and a Navarro-Frenk-White
(NFW; \citealt{Navarro_1996}) dark matter halo. The intrinsic stellar
mass distribution is found by deprojecting the surface brightness
model in Section \ref{sec:mge} assuming an oblate axisymmetric shape,
an inclination angle of 75$^\circ$, and a $V$-band stellar
mass-to-light ratio ($\Upsilon_\mathrm{V}$) that remains constant with
radius. NGC 1277 appears very flattened, is a fast rotator, and
doesn't show signs of kinematic twists -- all of which support the
assumption of axisymmetry. The inclination angle is the same as the
one assumed in previous studies of NGC 1277 \citep{vandenBosch_2008,
  Emsellem_2013, Yildirim_2015}, and is set by the nuclear dust disk
in the optical \emph{HST} image. In our models, the NFW halo is
described by the concentration parameter ($c$), which we set to 10,
and the fraction of dark matter relative to the stellar mass
($f_\mathrm{DM}$), which we set to values of 10, 100, and 1000,
corresponding to halo virial masses of $\sim$$10^{12}$,
$\sim$$10^{13}$, and $\sim$$10^{14}\ M_\odot$. We also ran models that
do not incorporate a dark halo. For $M_\mathrm{BH}$ and
$\Upsilon_\mathrm{V}$, we sampled 31 values between $5\times10^8$ and
$5\times10^{10}\ M_\odot$ and 41 values from $5.0-15.0\
\Upsilon_\odot$, respectively.

For a given galaxy potential set by $M_\mathrm{BH}$,
$\Upsilon_\mathrm{V}$, $c$, and $f_\mathrm{DM}$, we generated an orbit
library that samples 32 logarithmically spaced equipotential shells at
radii from 0\farcs003 to 85\arcsec, starting at 9 angular and 9 radial
values at each energy. We further bundle together 125 orbits with
adjacent initial conditions to ensure a smooth distribution function,
yielding a total of 972,000 orbits. The orbits are integrated in the
potential, and after accounting for the PSF and aperture binning,
weights are assigned to the orbits through a non-negative least
squares process such that their combined properties match the
intrinsic and projected stellar masses along with the observed
kinematics. We required that the surface brightness and
three-dimensional mass distributions be fit to an accuracy of
1\%. Since we are most interested in the mass of the black hole in NGC
1277, we reduce the effects on $M_\mathrm{BH}$ associated with our
choices of dark halo parameterization and a constant mass-to-light
ratio by fitting to only the NIFS kinematics. With four Gauss-Hermite
moments measured in 105 NIFS spatial bins, there are 420 observables.

\section{Results}
\label{sec:results}

\begin{figure*}
\begin{center}
\epsscale{1.1}
\plotone{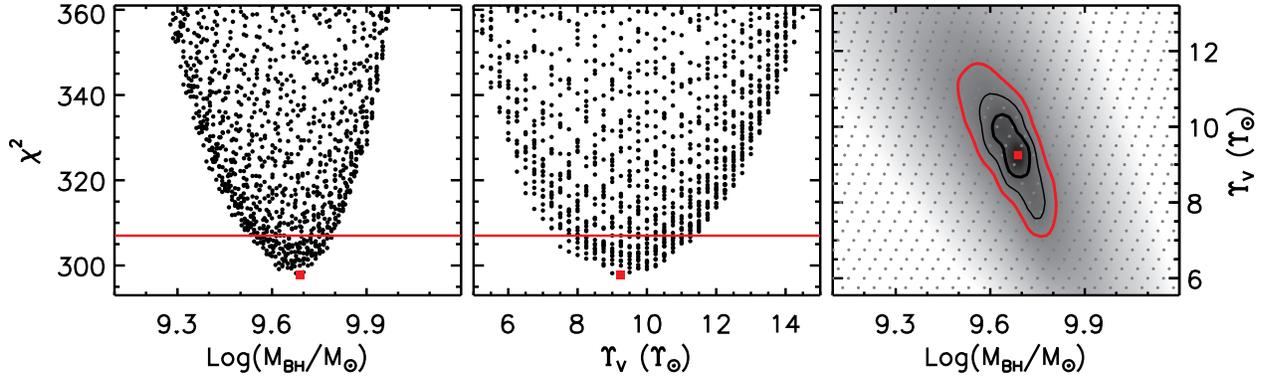}
\caption{Results of stellar-dynamical models fit to the NIFS
  kinematics. The $\chi^2$ is shown with black hole mass (left panel)
  and $V$-band mass-to-light ratio (middle panel). Each black dot
  corresponds to a single model, with the best-fit model highlighted
  as the red square. The red solid line indicates where $\Delta \chi^2
  = 9.0$, corresponding to the 3$\sigma$ confidence level for one
  parameter. $\chi^2$ contours are shown for a grid of black hole
  masses and mass-to-light ratios (right panel). Each gray point
  represents a model, and the best-fit model (red square) corresponds
  to $M_\mathrm{BH} = 4.9 \times 10^9\ M_\odot$ and
  $\Upsilon_\mathrm{V} = 9.3\ \Upsilon_\odot$. Relative to the
  minimum, the $\chi^2$ has increased by 2.3 (thick black contour),
  6.2 (thin black contour), and 11.8 (red contour), corresponding to
  the 1$\sigma$, 2$\sigma$, and 3$\sigma$ confidence intervals for two
  parameters. \label{fig:chi2_contour_nonly}}
\end{center}
\end{figure*}

\begin{figure*}
\begin{center}
\epsscale{0.9}
\plotone{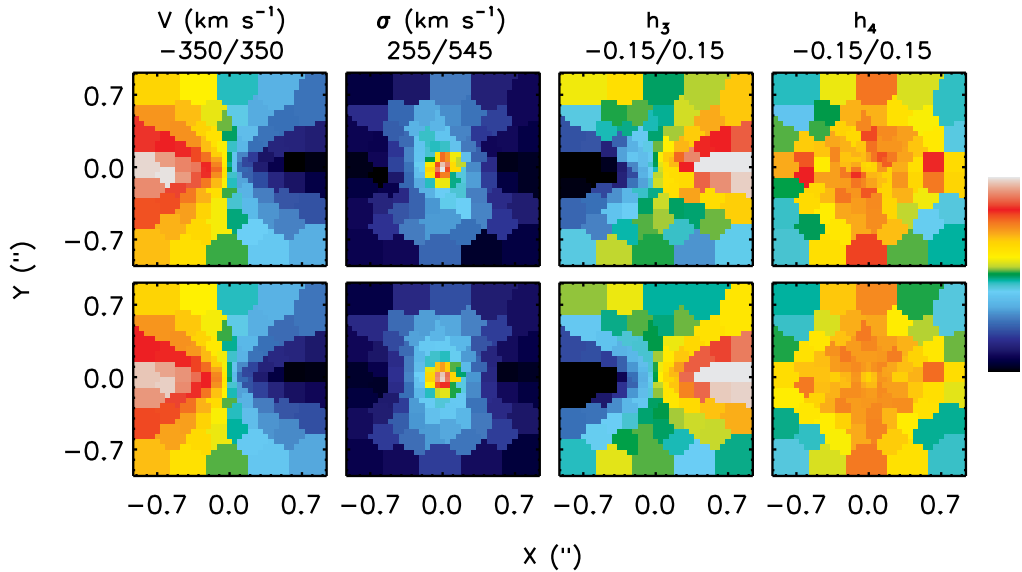}
\caption{The observed NIFS kinematics (top) show that NGC 1277 is
  rotating, with the west side of the galaxy being blue-shifted, that
  there is a sharp rise in the velocity dispersion and generally
  positive $h_4$ values at the nucleus, and that $h_3$ and $V$ are
  anti-correlated. The best-fit stellar dynamical model (bottom) is
  shown on the same scale given by the color bar to the right and the
  minimum/maximum values are provided at the top of the maps. The
  best-fit model nicely reproduces the NIFS observations and has a
  reduced $\chi^2$ of $0.7$. \label{fig:nifsbestmodel_nonly}}
\end{center}
\end{figure*}

When fitting stellar-dynamical models to the point-symmetrized NIFS
kinematics, we find that $M_\mathrm{BH}$ is well constrained, as can
be seen in Figure \ref{fig:chi2_contour_nonly}. We measured the formal
1$\sigma$ and 3$\sigma$ modeling fitting uncertainties by examining
the $\chi^2$ after marginalizing over the other three parameters and
searching for where the $\chi^2$ has increased by $1$ and $9$,
respectively, from the minimum ($\chi^2_\mathrm{min}$). This
corresponds to $M_\mathrm{BH} = (4.9\pm0.2) \times 10^9\ M_\odot$ and
$\Upsilon_\mathrm{V} = 9.3\pm0.4\ \Upsilon_\odot$ (1$\sigma$), and
$M_\mathrm{BH} = (4.9^{+1.2}_{-1.4}) \times 10^9\ M_\odot$ and
$\Upsilon_\mathrm{V} = 9.3^{+2.1}_{-1.9}\ \Upsilon_\odot$
(3$\sigma$). As expected, the dark halo cannot be constrained with the
NIFS data; all three dark halos and the model without dark matter are
within $\Delta \chi^2 \equiv \chi^2 - \chi^2_\mathrm{min} = 5$. The
best-fit model is compared to the observed NIFS kinematics in Figure
\ref{fig:nifsbestmodel_nonly} and Figure \ref{fig:nifsvshetvsppak},
and is an excellent match to the data with a reduced $\chi^2$ of
0.7. The best-fit model is able to reproduce the dramatic rise in the
stellar velocity dispersion at the nucleus and the slight increase in
$h_4$ at the center.

We further tested the robustness of the $M_\mathrm{BH}$ and
$\Upsilon_\mathrm{V}$ results by examining the change in the best-fit
values when assuming a different PSF, when using unsymmetrized
kinematics, and when adopting a different pPXF continuum-correction
model while extracting the NIFS kinematics. Given the challenges
associated with the PSF measurement due to the evolving AO correction
and the merging of data cubes that span several nights, it is
important to test how other reasonable PSF parameterizations affect
$M_\mathrm{BH}$. For the fiducial model presented above, we utilized a
two-Gaussian PSF model measured from the comparison of the MGE to the
NGC 1277 collapsed NIFS data cube. As a secondary approach, we test
using the sum of four Gaussians to describe the PSF, which was
determined from NIFS PSF star observations. In comparison to the
two-Gaussian PSF model, the core of the four-Gaussian PSF model has a
similar dispersion but a lower weight, such that the core makes up
45\% of the total flux. Using this four-Gaussian PSF model results in
the best-fit $M_\mathrm{BH}$ increasing by 22\% to $6.0 \times 10^9\
M_\odot$ and $\Upsilon_\mathrm{V}$ decreasing by 14\% to $8.0\
\Upsilon_\odot$.

The next test we ran involved fitting stellar-dynamical models to the
unsymmetrized NIFS kinematics. Symmerization is commonly used to
reduce noise in the observed kinematic measurements. Our fiducial
model was fit to NIFS kinematics that were averaged in a two-fold
symmetric way about the major and minor axes while accounting for the
flux each spaxel contributed to the spatial bin and the kinematic
error of the bin. If instead no adjustments are made to the observed
NIFS kinematics, with the exception of subtracting off the offsets in
the odd Gauss-Hermite moments (e.g., the galaxy's recessional
velocity), then we measure a black hole mass of $5.8 \times 10^9\
M_\odot$ and a $V$-band stellar mass-to-light ratio of $8.8\
\Upsilon_\odot$. Hence, $M_\mathrm{BH}$ increases by 18\% and
$\Upsilon_\mathrm{V}$ decreases by 5\%. The best-fit model to the
unsymmetrized NIFS kinematics has a reduced $\chi^2$ of 2.9.

Finally, the NIFS kinematics appear to be sensitive to the adopted
pPXF continuum-correction model. We measured the stellar kinematics by
comparing spectra of K and M giant and supergiant stars to the NGC
1277 spectra. Since the stars were also observed with NIFS, their
spectra have a similar shape as the galaxy spectra; however, minor
adjustments to the templates are needed to account for differences in
the CO bandhead equivalent widths and for reddening/imperfect flux
calibration. In order to provide a better match to the observed galaxy
spectra, we used additive/multiplicative Legendre polynomials with
pPXF. As described in Section \ref{sec:nifskin}, we selected an
additive degree 0 and multiplicative degree 1 polynomial, and allowed
pPXF to solve for the best-fit coefficients simultaneously with the
Gauss-Hermite moments. An additive degree 0 and multiplicative degree
1 polynomial produced a good fit to the NGC 1277 observations, and the
use of other low-order polynomials yield consistent kinematics.

Adopting higher-order polynomials resulted in slightly better fits to
the galaxy spectra, but generally lower stellar velocity dispersions,
especially at the nucleus. Notably, extracting the NIFS kinematics
using a multiplicative degree 2 polynomial resulted in velocity
dispersions that were lower by as much as 36 km s$^{-1}$ compared to
when an additive degree 0 with a multiplicative degree 1 polynomial
were utilized. Other combinations of multiplicative degree $2-4$
polynomials with no additive contribution, or with additive degree
$0-1$ polynomials, produced consistent kinematics as the
multiplicative degree 2 polynomial case. Therefore, we tested the
effect on $M_\mathrm{BH}$ when fitting to point-symmetrized NIFS
kinematics extracted with an order 2 multiplicative polynomial. We
found that the black hole mass decreased to $4.2 \times 10^9\
M_\odot$, or by 14\%, and the $V$-band stellar mass-to-light ratio
increased to $10.0\ \Upsilon_\odot$, or by 8\%.

The final uncertainties are determined by adding in quadrature the
change in the $M_\mathrm{BH}$ and $\Upsilon_\mathrm{V}$ from each of
the three tests above and the 1$\sigma$ statistical errors from the
fiducial model grid. This results in fractional errors of 32\% and
17\% on $M_\mathrm{BH}$ and $\Upsilon_\mathrm{V}$, respectively. In
other words, $M_\mathrm{BH} = (4.9 \pm 1.6) \times 10^9\ M_\odot$ and
$\Upsilon_\mathrm{V} = 9.3 \pm 1.6\ \Upsilon_\odot$. The systematics
associated with the NIFS kinematics and NIFS PSF input into the
stellar-dynamical models dominate over the formal 1$\sigma$ model
fitting uncertainty. In terms of other common systematic sources of
error that plague stellar-dynamical measurements, by fitting to just
the NIFS kinematics we have limited mass degeneracies with the
galaxy's dark halo \citep[e.g.,][]{Gebhardt_Thomas_2009, Rusli_2013}
and uncertainties associated with the assumption of a constant stellar
mass-to-light ratio \citep[e.g.,][]{McConnell_2013}. Moreover, owing
to the presence of the nuclear dust disk, uncertainty in the viewing
orientation is minimal, unlike some past stellar-dynamical studies
\citep[e.g.,][]{Shapiro_2006, Onken_2007}. The dust disk, however, may
be problematic for construction of the luminous mass model in Section
\ref{sec:mge}. We conservatively masked the dust when generating the
MGE model, but acknowledge that if the MGE under-predicts the nuclear
stellar mass then the black hole mass would be overestimated.

\subsection{Models with Large-scale Kinematics}
\label{sec:largescaledata}

\begin{figure*}
\begin{center}
\epsscale{0.95}
\plotone{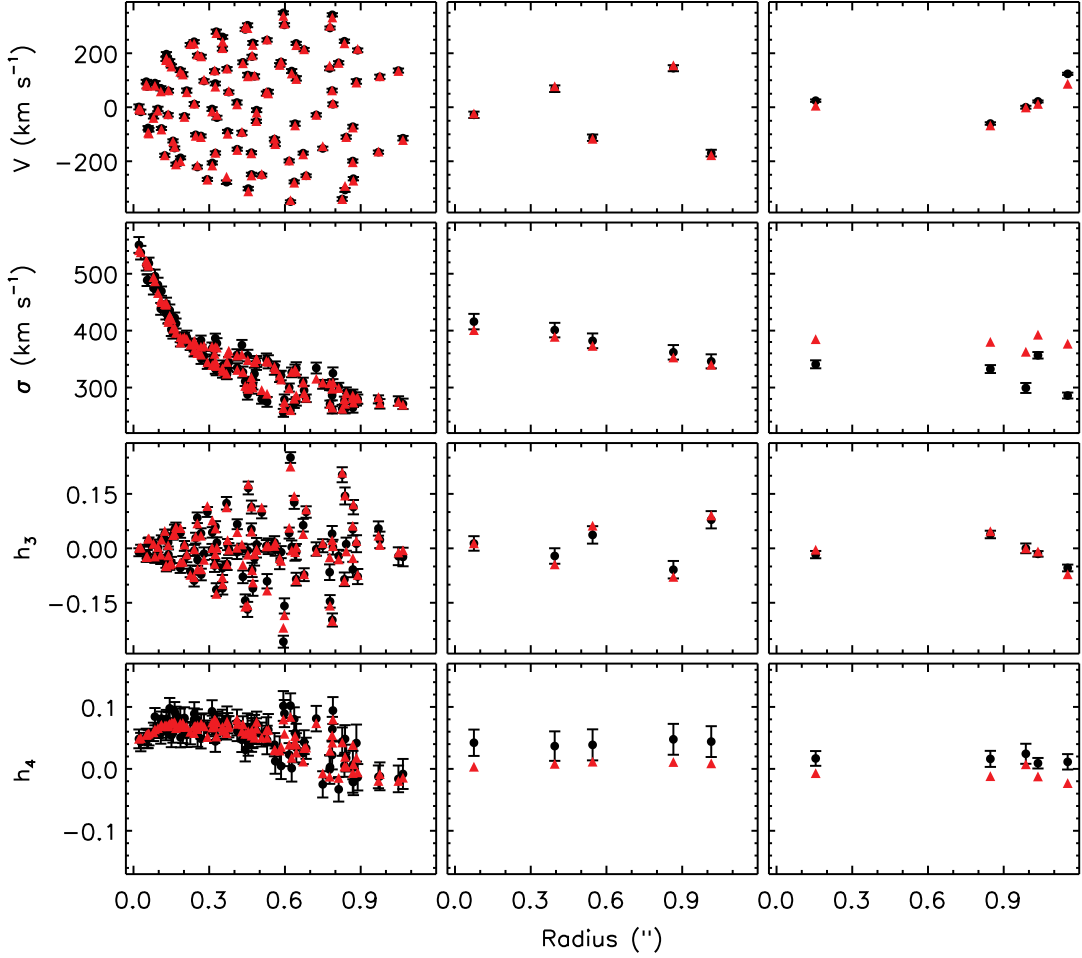}
\caption{The observed kinematics (black circles) measured from NIFS
  (left column), HET (middle column), and PPAK (right column) are
  shown over the radial extent of the NIFS data. The data have been
  folded and are plotted as a function of projected distance from the
  nucleus. While the HET measurements have been made along the galaxy
  major axis, multiple position angles are depicted for the IFU
  data. For comparison, we plot the best-fit model constrained by only
  the NIFS kinematics (red triangles). Thus, the best-fit model
  predictions for the central HET and PPAK kinematics take into
  account differences in spatial resolution. The best-fit model
  generally reproduces the observed central HET kinematics but is a
  worse match to the observed innermost PPAK kinematics. Instead,
  fitting dynamical models to NIFS+HET and to NIFS+PPAK produces black
  hole masses and mass-to-light ratios that are consistent with
  $M_\mathrm{BH} = (4.9\pm1.6) \times 10^9\ M_\odot$ and
  $\Upsilon_\mathrm{V} = 9.3\pm1.6\ \Upsilon_\odot$, found by fitting
  models to NIFS only. \label{fig:nifsvshetvsppak}}
\end{center}
\end{figure*}

Large-scale data is commonly used to complement high angular
resolution spectroscopy in order to better constrain the stellar
mass-to-light ratio and orbital distribution
\citep[e.g.,][]{Shapiro_2006, Krajnovic_2009}. The NIFS kinematics
extend out to a radius of 1\farcs3 (442 pc), or $\sim$0.4
$R_\mathrm{e}$. Therefore, the NIFS data not only resolve the black
hole sphere of influence, but also probe the region where the stars
begin to dominate the potential. It is encouraging that our fiducial
model presented at the beginning of Section \ref{sec:results} largely
reproduces the observed HET kinematics over the radial range of the
NIFS data, as can be seen in Figure \ref{fig:nifsvshetvsppak}. The
predictions from this best-fit model, constrained by only the NIFS
kinematics, are consistent with the velocity, velocity dispersion, and
$h_3$ HET measurements, and only deviate from the central $h_4$ HET
values by about 1.5 times the measurement uncertainty. Although the
fiducial model is in fair agreement with the HET observations, Figure
\ref{fig:nifsvshetvsppak} shows that the same model is worse at
matching the observed PPAK kinematics. The predictions from the
best-fit model constrained only by the NIFS data are consistent with
the velocity and $h_3$ PPAK measurements, but over-estimate the
central PPAK velocity dispersions by about nine times the measurement
errors, in addition to under-predicting the inner $h_4$ PPAK values by
twice the uncertainty.

Therefore, we also ran dynamical models that fit to the combination of
NIFS and HET kinematics, as well as to the combination of NIFS and
PPAK kinematics, continuing to sample the same dark halos described in
Section \ref{sec:modeling}. We recovered best-fit values of
$M_\mathrm{BH} = 4.8 \times 10^9\ M_\odot$ and $\Upsilon_\mathrm{V} =
10.3\ \Upsilon_\odot$ with a reduced $\chi^2$ of 0.8 when fitting all
31 HET spatial bins, and $M_\mathrm{BH} = 4.6 \times 10^9\ M_\odot$
and $\Upsilon_\mathrm{V} = 9.8\ \Upsilon_\odot$ with a reduced
$\chi^2$ of 0.8 when excluding three outer bins because their
dispersions are below the HET/LRS instrumental resolution. Instead if
dynamical models are fit to the combination of the NIFS and PPAK
kinematics, we infer $M_\mathrm{BH} = 6.1 \times 10^9\ M_\odot$ and
$\Upsilon_\mathrm{V} = 8.3\ \Upsilon_\odot$ with a reduced $\chi^2$ of
0.8. In summary, we find that the black hole mass and mass-to-light
ratio measured from models fit to NIFS+HET and NIFS+PPAK agree with
the results from our fiducial model, fit to NIFS only, after
accounting for systematics associated with the NIFS kinematics and
PSF.

\section{Discussion}
\label{sec:discussion}

Below we compare our black hole mass measurement to the previous mass
determinations in the literature and discuss the implications of NGC
1277's location on the black hole$-$host galaxy relations.

\subsection{Comparison to Previous Black Hole Mass Measurements}
\label{subsec:mbh_comparison}

With $M_\mathrm{BH} = 4.9 \times 10^9\ M_\odot$ and $\sigma_\star =
333$ km s$^{-1}$, the NIFS observations probe well within the black
hole sphere of influence of 0\farcs6. Previous stellar-dynamical work
using seeing-limited data from long-slit HET observations and IFU PPAK
observations found $M_\mathrm{BH} = 1.7 \times 10^{10}\ M_\odot$
\citep{vandenBosch_2012} and $M_\mathrm{BH} = 1.2 \times 10^{10}\
M_\odot$ \citep{Yildirim_2015}, respectively. Neither study, however,
quantified the impact of possible systematic effects on the NGC 1277
black hole mass.

In particular, \cite{Emsellem_2013} found that the HET kinematics
presented in \cite{vandenBosch_2012} could be nearly reproduced with a
$5 \times 10^9\ M_\odot$ black hole. A rather small deviation between
the $5 \times 10^9\ M_\odot$ black hole model and the HET kinematics
occurs for the innermost $h_4$ measurements, such that the model
under-predicts the observed values by about twice the uncertainty,
which is similar to our best-fit model. Given the possibility of
spurious kinematic measurements due to issues like template mismatch,
\cite{Emsellem_2013} advocated for the use of 3$\sigma$ confidence
levels for a more conservative estimate of the $M_\mathrm{BH}$
uncertainty. For stellar-dynamical models fit to the HET kinematics,
the 3$\sigma$ uncertainty corresponds to $M_\mathrm{BH} = (6.5 - 32)
\times 10^9\ M_\odot$ and $\Upsilon_\mathrm{V} = 0.1 - 9.7\
\Upsilon_\odot$ \citep{vandenBosch_2012}.

Similarly, \cite{Yildirim_2015} opted to report 3$\sigma$ statistical
uncertainties as a careful way in which to assign errors to
$M_\mathrm{BH}$ and $\Upsilon_\mathrm{V}$. They fit to both PPAK and
HET observations simultaneously to determine $M_\mathrm{BH} =
(1.3^{+0.3}_{-0.5}) \times 10^{10}\ M_\odot$ and $\Upsilon_\mathrm{V}
= 6.5 \pm 1.5\ \Upsilon_\odot$. They comment though that the central
dispersion and $h_4$ measurements from the PPAK data are
systematically lower than the inner values from the HET data, which
could be due to a number of reasons (e.g., the use of different
stellar libraries, different spectral fitting regions, different
spatial resolutions, etc.). Since the PPAK data provide an independent
check of the validity of the HET kinematics, \cite{Yildirim_2015}
additionally fit orbit-based models to only the PPAK kinematics,
finding a lower limit of $M_\mathrm{BH} = 4.0 \times 10^9\ M_\odot$,
which they view as a very conservative bound on the NGC 1277 black
hole mass.

Therefore, our black hole mass of $(4.9 \pm 1.6) \times 10^9\ M_\odot$
and $V$-band mass-to-light ratio of $9.3 \pm 1.6\ \Upsilon_\odot$ are
in agreement with the above conservative limits from
\cite{vandenBosch_2012} and \cite{Yildirim_2015}. Although it is
tempting to assume that the smaller black hole mass inferred from the
NIFS data is the result of smaller central $h_4$ values as suggested
by Figure \ref{fig:nifsvshetvsppak}, much of the inferred black hole
mass is being driven by the lower order moments, and not $h_4$. If
dynamical models are fit to only the NIFS velocity, velocity
dispersion, and $h_3$ measurements, we recover a best-fit black hole
mass consistent with $M_\mathrm{BH} = (4.9 \pm 1.6) \times 10^9\
M_\odot$. Similarly, if an artificial constant is added to the
observed NIFS $h_4$ values (we tested adding an offset of 0.03 and
0.06), no change in the black hole mass outside of the quoted
uncertainties is observed.

A more rigorous comparison ultimately requires an attempt to quantify
some possible systematics that affect $M_\mathrm{BH}$ and
$\Upsilon_\mathrm{V}$ derived from the low spatial resolution
data. For example, the HET and PPAK kinematics from
\cite{vandenBosch_2012} and \cite{Yildirim_2015} extend to $\sim$3
$R_\mathrm{e}$ ($\sim$10\arcsec), and both studies assumed that the
stellar mass-to-light ratio remains constant over this radial
range. \cite{Yildirim_2015} argued against a varying mass-to-light
ratio based on a lack of color gradients in Sloan Digital Sky Survey
(SDSS) imaging, but \cite{MartinNavarro_2015} find a possible increase
in $\Upsilon_\mathrm{V}$ from $7.5\ \Upsilon_\odot$ at $\sim$5\arcsec\
to $11.6\ \Upsilon_\odot$ at $\sim$1\arcsec\ by measuring gravity
sensitive absorption lines in optical/near-infrared spectra of NGC
1277 and assuming a bimodal initial mass function. Since the
large-scale HET and PPAK spectroscopy do not resolve the black hole
sphere of influence, the assumptions made about the mass-to-light
ratio will have a significant impact on $M_\mathrm{BH}$ (e.g.,
\citealt{Rusli_2013}). If $\Upsilon_\mathrm{V}$ is actually larger at
the center than that assumed by \cite{vandenBosch_2012} and
\cite{Yildirim_2015}, then their inferred $M_\mathrm{BH}$ is
overestimated.

The factor of $\sim$3 difference in the best-fit black hole masses
between this work and the prior stellar-dynamical studies is
large. However, incorporating any identifiable systematics into the
error budget may make the $M_\mathrm{BH}$ and $\Upsilon_\mathrm{V}$
measurements from the AO spectroscopy and the large-scale spectroscopy
come into better agreement than it currently appears based simply on
the 1$\sigma$ model fitting uncertainties. We have done this in
Section \ref{sec:results} for $M_\mathrm{BH}$ and
$\Upsilon_\mathrm{V}$ derived from the AO observations; it is notable
that the uncertainties associated with the extraction of the NIFS
kinematics and the NIFS PSF alone dominate over the 1$\sigma$
statistical errors.

In addition to above stellar-dynamical black hole mass measurements,
recently \cite{Scharwachter_2015} presented cold molecular gas
kinematics from low angular resolution radio observations of NGC
1277. \cite{Scharwachter_2015} modeled the molecular gas distribution
from the $\sim$1\arcsec\ resolution and $\sim$2\farcs5 resolution
configurations of PdBI as a nuclear ring and as an exponential disk
with an inclination angle of $75^\circ$. Since the CO(1-0) emission
was spatially unresolved, the black hole mass was not well constrained
and they found that the double-horned CO profile was consistent with
both $M_\mathrm{BH} = 5 \times 10^9\ M_\odot$ and $M_\mathrm{BH} = 1.7
\times 10^{10}\ M_\odot$. Thus, the black hole mass measurement
presented here is consistent with constraints from molecular gas
kinematics, and it would be interesting to compare to higher angular
resolution gas kinematics. Currently, there are a very limited number
of meaningful comparisons between stellar and gas-dynamical black hole
mass measurement methods (e.g., \citealt{deFrancesco_2006,
  Neumayer_2007, Cappellari_2009, Gebhardt_2011, Walsh_2012,
  Walsh_2013}).

\subsection{NGC 1277 on the Black Hole Scaling Relations}
\label{subsec:scalingrelations}

\begin{figure*}
\begin{center}
\epsscale{1.0}
\plotone{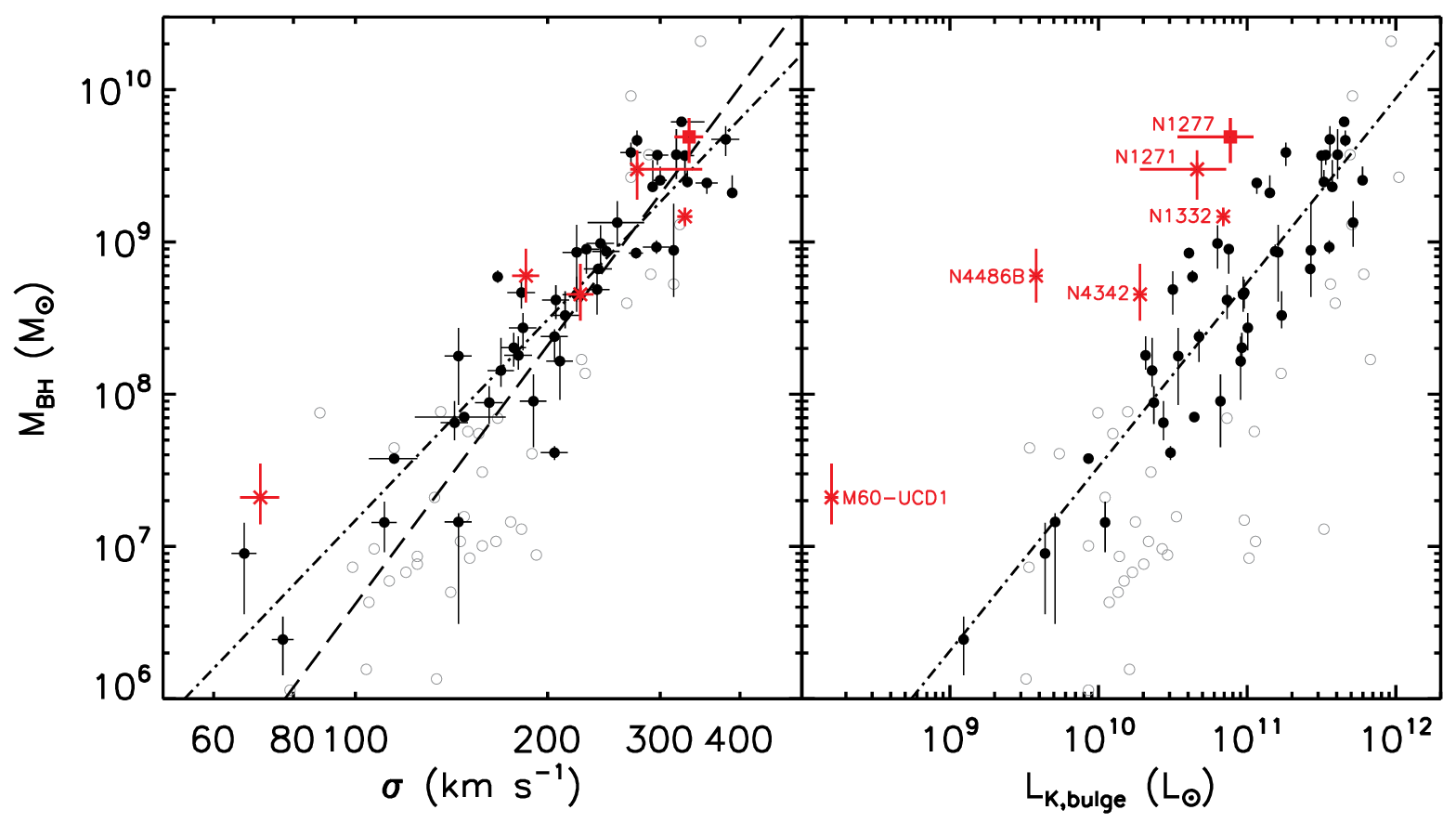}
\caption{The location of NGC 1277 (red square) and other similar
  galaxies (red asterisks) on the black hole$-$host galaxy
  correlations. The $\sim$80 galaxies with dynamical black hole mass
  measurements shown here are taken from \cite{Kormendy_Ho_2013}, with
  the exception of M60-UCD1 \citep{Seth_2014}, NGC 1332
  \citep{Rusli_2011}, NGC 1271 \citep{Walsh_2015}, and NGC 1277 (this
  work). The dot-dashed lines show the fitted relations from
  \cite{Kormendy_Ho_2013} to the black points. The gray points,
  composed largely of galaxies with pseudo-bulges, were excluded from
  the \cite{Kormendy_Ho_2013} fit. For comparison, the steeper dashed
  line is the relation from \cite{McConnell_Ma_2013}, which was fit to
  nearly all points. NGC 1277 and the galaxies labeled in red have
  small sizes and large velocity dispersions for their galaxy
  luminosities, and they harbor over-massive black holes relative to
  $M_\mathrm{BH} - L_\mathrm{bul}$ but are consistent with the
  $M_\mathrm{BH} - \sigma_\star$ relation. \label{fig:mbhrels}}
\end{center}
\end{figure*}

\begin{figure*}
\begin{center}
\epsscale{0.9}
\plotone{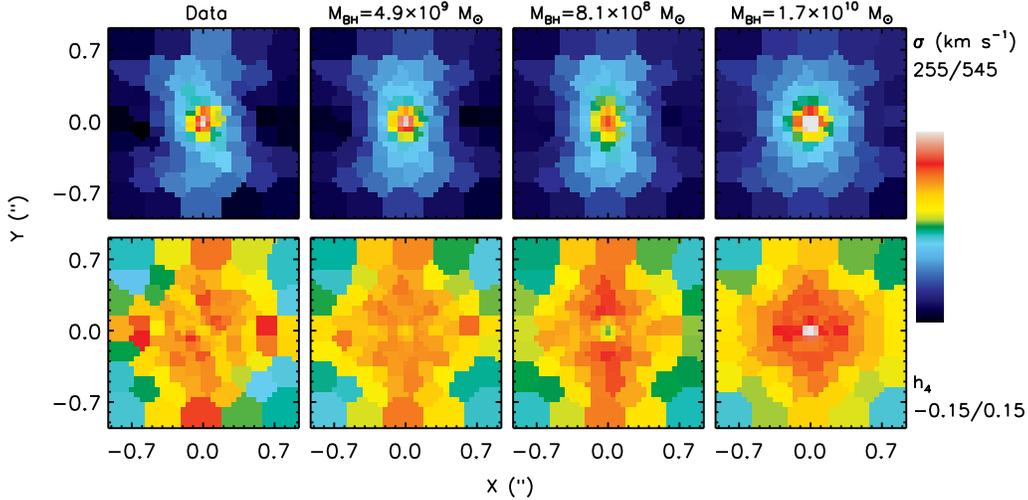}
\caption{Comparison between the NIFS velocity dispersion and $h_4$
  (left) and the best-fit model with $M_\mathrm{BH} = 4.9 \times 10^9\
  M_\odot$ (middle left), a model with $M_\mathrm{BH} = 8.1 \times
  10^8\ M_\odot$ (middle right), and a model with $M_\mathrm{BH} = 1.7
  \times 10^{10}\ M_\odot$ (right). A mass of $8.1 \times 10^8\
  M_\odot$ is expected based on the $M_\mathrm{BH} - L_\mathrm{bul}$
  relation when conservatively using the galaxy's total $K$-band
  luminosity of $1.4 \times 10^{11}\ L_\odot$, whereas a black hole
  mass of $1.7 \times 10^{10}\ M_\odot$ is the previously published
  result from stellar-dynamical modeling of seeing-limited,
  large-scale spectroscopy \citep{vandenBosch_2012}. When constructing
  the $8.1 \times 10^8\ M_\odot$ black hole model, we sample over
  mass-to-light ratios and dark halos as described in Section
  \ref{sec:modeling}, and present the model with the lowest $\chi^2$
  for this black hole mass. The $1.7 \times 10^{10}\ M_\odot$ black
  hole model uses the same mass-to-light ratio and dark halo found by
  \cite{vandenBosch_2012}. The best-fit model matches the observed
  velocity dispersion peak and slightly positive nuclear $h_4$ values
  very well, however the smaller (larger) black hole under-predicts
  (over-predicts) the central velocity dispersion and $h_4$
  values. \label{fig:nifscomparemodels}}
\end{center}
\end{figure*}

The smaller inferred black hole mass from this work does not change
the main result that NGC 1277 harbors an over-massive black hole for
the galaxy's luminosity, although NGC 1277 is now more consistent with
the $M_\mathrm{BH} - \sigma_\star$ relation. In Figure
\ref{fig:mbhrels}, we plot NGC 1277 on the black hole-galaxy
relations. The black hole mass expected from $M_\mathrm{BH} -
L_\mathrm{bul}$ is $(1.5-6.1)\times10^8\ M_\odot$ for $K$-band bulge
luminosities between $(3.4- 11.0)\times10^{10}\ L_\odot$
\citep{Kormendy_Ho_2013}, while the mass expected from $M_\mathrm{BH}
- \sigma_\star$ is $(2.9-3.7) \times 10^9\ M_\odot$ depending on
whether the \cite{McConnell_Ma_2013} or \cite{Kormendy_Ho_2013}
calibration is used. Therefore, NGC 1277 is an order of magnitude
above the mass expected from the $M_\mathrm{BH} - L_\mathrm{bul}$
relation. The model with a $8.1 \times 10^8\ M_\odot$ black hole,
which is the mass expected from $M_\mathrm{BH} - L_\mathrm{bul}$ when
conservatively using the galaxy's total luminosity, is clearly a poor
match to the NIFS observations and under-predicts the strong rise in
the nuclear velocity dispersion and the minor increase in central
$h_4$ values, as can be seen in Figure \ref{fig:nifscomparemodels}. In
contrast, the model with a larger black hole of $M_\mathrm{BH} = 1.7
\times 10^{10}\ M_\odot$, which is the best-fit mass from
\cite{vandenBosch_2012}, clearly over-estimates the NIFS central
velocity dispersion and $h_4$ values.

NGC 1277 is similar to the galaxies highlighted in
\cite{Kormendy_Ho_2013} as ``black hole monsters'', namely NGC 4486B,
NGC 4342, and NGC 1332, as well as the recently studied galaxies
M60-UCD1 and NGC 1271. In particular, NGC 4342, NGC 1332, and NGC 1271
appear most like NGC 1277. All of these early-type galaxies are small,
are outliers from the Faber-Jackson relation (i.e., show high velocity
dispersions for their luminosities), are rotating, and are positively
offset from $M_\mathrm{BH} - L_\mathrm{bul}$ but consistent with
$M_\mathrm{BH} - \sigma_\star$. Although less certain, there are hints
that the galaxies Mrk 1216 \citep{Yildirim_2015} and SDSS
J151741.75-004217.6 \citep{Lasker_2013} are also offset from
$M_\mathrm{BH} - L_\mathrm{bul}$. We note that this behavior could
differ from that observed for BCGs, as new measurements
\citep{McConnell_2012} may indicate that the $M_\mathrm{BH} -
\sigma_\star$ relationship saturates, such that $M_\mathrm{BH}$
becomes independent of $\sigma_\star$ at large masses, but that BCGs
continue to follow $M_\mathrm{BH} - L_\mathrm{bul}$
\citep{McConnell_Ma_2013, Kormendy_Ho_2013}. If compact,
high-dispersion galaxies like NGC 1277 and BCGs indeed exhibit
different scaling relations, that would suggest that the black
holes/galaxies grew via different pathways.

There are a few potential explanations for the presence of these
over-massive black holes. One interesting possibility is that the
galaxies are relics of the $z\sim2$ quiescent galaxies, passively
evolving since that epoch. \cite{FerreMateu_2015} argue this is the
case for NGC 1277 and NGC 1271, citing uniform old stellar populations
(ages $\gtrsim$10 Gyr) and small sizes that make the galaxies outliers
from the local mass-size relation of \cite{Shen_2003}. An equivalent
picture was previously discussed by \cite{Trujillo_2014}, who also
find uniform old stellar ages for NGC 1277, make note of the galaxy's
massive and compact nature, and point to an identical stellar mass
density to the massive, high-$z$ passive galaxies as another
indication that NGC 1277 is a relic galaxy. In particular,
\cite{FerreMateu_2015} and \cite{Trujillo_2014} suggest that the
compact galaxies took a different evolutionary path than the one
assumed for local massive galaxies, skipping a phase in which galaxies
grow in size and mildly in mass via mergers after $z\sim$2. Most
recently, \cite{Wellons_2015} echoed an analogous sentiment based on
the analysis of Illustris simulations. Likewise, \cite{Fabian_2013}
proposed that a majority of black hole growth occurred before $z\sim$3
and that the accretion of gas and stars dictates whether there is
subsequent evolution of the compact bulges. In other words, maybe
galaxies like NGC 1277 are positive outliers from the local
$M_\mathrm{BH} - L_\mathrm{bul}$ relation because they reflect a
previous time when galaxies harbored over-massive black holes, and the
growth of host galaxies had yet to catch up.

A second explanation for the over-massive black holes is that these
objects are simply unusual, and are found in the tails of a
distribution between black hole masses and host galaxy
properties. With the limited number of dynamical measurements for
$\gtrsim$10$^9\ M_\odot$ black holes currently available, the form and
scatter of the relations in this high-mass regime are not well
established \citep{McConnell_Ma_2013}. Therefore, it is difficult to
assess how much these compact, high-dispersion galaxies in reality
deviate from the local $M_\mathrm{BH} - L_\mathrm{bul}$
relation. Another way in which to create an over-massive black hole is
through tidal stripping. This is believed to the be the case for
M60-UCD1 \citep{Seth_2014}, however NGC 1277 does not show signs of
recent interactions and the isophotes from the \emph{HST} image appear
regular without asymmetries. Another idea was put forth by
\cite{Shields_Bonning_2013}, who examine the possibility that the
black hole was ejected from the BCG of Perseus, wandered about the
cluster core, and was subsequently captured by NGC 1277. This,
however, requires a series of rare events, and the scenario has become
more unlikely as we now have confirmed another over-massive black hole
\citep{Walsh_2015} and have a number of other candidates
\citep{vandenBosch_2012, vandenBosch_2015, FerreMateu_2015} from
galaxies also located in the Perseus cluster.

More dynamical mass measurements of the black holes in galaxies like
NGC 1277 are needed, especially if such objects turn out to be the
relics of higher redshift quiescent galaxies. Roughly 1 in 100 of the
$z$$\sim$2 passive galaxies survive until today \citep{Trujillo_2009,
  vanderWel_2014, Saulder_2015}. Although such relics are thought to
be rare and are not representative of the local galaxy population,
dynamically measuring $M_\mathrm{BH}$ in a significant sample will
provide insight into the growth of black holes with their host
galaxies over time. The HET Massive Galaxy Survey
\citep{vandenBosch_2015} has uncovered about a dozen galaxies with
sizes, dispersions, and luminosities similar to NGC 1277. We currently
have high angular resolution imaging and spectroscopy for half of
these objects, and are working on acquiring secure $M_\mathrm{BH}$
measurements.

\section{Conclusion}
\label{sec:conclusion}

We have presented $K$-band spectroscopic observations of the nearby S0
galaxy NGC 1277 obtained with Gemini/NIFS assisted by AO. The
observations allow us to map out the stellar kinematics well with the
black hole's gravitational sphere of influence, and when combined with
high resolution \emph{HST} imaging, we are able to place strong
constraints on the mass of the central black hole. We find that
$M_\mathrm{BH} = (4.9 \pm 1.6) \times 10^9\ M_\odot$. These error bars
reflect the 1$\sigma$ statistical uncertainties in combination with
systematics associated with the NIFS kinematic measurements and the
PSF model. We elect to fit orbit-based dynamical models to only the
NIFS kinematics in order to limit other common systematic effects
known to affect black hole mass measurements; namely degeneracies
between the dark halo/the mass-to-light ratio/the black hole mass and
the assumption of a stellar mass-to-light ratio that remains constant
with radius. We note, however, that fitting dynamical models to the
NIFS kinematics in combination with the large-scale HET kinematics,
and in combination with the wide-field IFU PPAK kinematics, produces a
black hole mass consistent with $M_\mathrm{BH} = (4.9 \pm 1.6) \times
10^9\ M_\odot$.

The black hole mass measurement presented here is a factor of $\sim$3
smaller than the best-fit masses previously reported by
\cite{vandenBosch_2012} and \cite{Yildirim_2015}, both of which were
determined from stellar-dynamical modeling of large-scale,
seeing-limited spectroscopy. Neither study explicitly explored the
effect of systematics on $M_\mathrm{BH}$, which could significantly
increase the range of plausible black holes masses. When instead
considering the conservative lower 3$\sigma$ limit of $6.5 \times
10^9\ M_\odot$ from \cite{vandenBosch_2012} and of $4.0 \times 10^9\
M_\odot$ from \cite{Yildirim_2015}, our result is consistent with the
previous studies. Furthermore, our mass measurement agrees with the
constraints found from rotating molecular
gas. \cite{Scharwachter_2015} presented unresolved CO(1-0) emission
arising from the nuclear dust disk in NGC 1277, and found that both a
$\sim$$5 \times 10^9\ M_\odot$ and a $\sim$$1.7 \times 10^{10}\
M_\odot$ black hole are capable of reproducing the double-horned
profile.

With $M_\mathrm{BH} = (4.9 \pm 1.6) \times 10^9\ M_\odot$, the black
hole in NGC 1277 is one of the largest detected to date. NGC 1277
remains a positive outlier on the $M_\mathrm{BH} - L_\mathrm{bul}$
relation, but is consistent with $M_\mathrm{BH} - \sigma_\star$. NGC
1277 is thought to be a relic of the $z$$\sim$2 massive quiescent
galaxies \citep{Trujillo_2014, FerreMateu_2015}, and could suggest
that galaxies contained over-massive black holes at earlier epochs,
such that the growth of black holes precedes the growth of their host
galaxies.

\acknowledgements

Based on observations obtained at the Gemini Observatory acquired
through the Gemini Science Archive and processed using the Gemini IRAF
package, which is operated by the Association of Universities for
Research in Astronomy, Inc., under a cooperative agreement with the
NSF on behalf of the Gemini partnership: the National Science
Foundation (United States), the National Research Council (Canada),
CONICYT (Chile), the Australian Research Council (Australia),
Minist\'{e}rio da Ci\^{e}ncia, Tecnologia e Inova\c{c}\~{a}o (Brazil)
and Ministerio de Ciencia, Tecnolog\'{i}a e Innovaci\'{o}n Productiva
(Argentina), under program GN-2011B-Q-27. Based on observations made
with the NASA/ESA Hubble Space Telescope, and obtained from the Hubble
Legacy Archive, which is a collaboration between the Space Telescope
Science Institute (STScI/NASA), the Space Telescope European
Coordinating Facility (ST-ECF/ESA) and the Canadian Astronomy Data
Centre (CADC/NRC/CSA). The authors acknowledge the Texas Advanced
Computing Center (TACC; http://www.tacc.utexas. edu) at the University
of Texas at Austin for providing HPC resources that have contributed
to the research results reported within this paper. This research has
made use of the NASA/IPAC Extragalactic Database which is operated by
the Jet Propulsion Laboratory, California Institute of Technology,
under contract with NASA.

\end{document}